%% file: ReactionPRC.tex
\documentclass[twocolumn,showpacs,preprintnumbers,amsmath,amssymb,prc,nofootinbib]{revtex4}

\usepackage{graphicx}
\usepackage{dcolumn}
\usepackage{bm}
\usepackage{multirow}
\usepackage{color}

\include{comdefs}

\begin{document}

\title{Towards a Microscopic Reaction Description Based on Energy-Density-Functional Structure Models}

\author{G. P. A. Nobre\footnote{Current address: Brookhaven National Laboratory, Building 197D, Upton, NY 11973-5000, USA;  \texttt{gnobre@bnl.gov}}}
\author{F. S. Dietrich}
\author{J. E. Escher}
\author{I. J. Thompson}
\affiliation{Lawrence Livermore National Laboratory, P. O. Box 808, L-414, Livermore, CA 94551, USA}
\author{M. Dupuis}
\affiliation{CEA, DAM, DIF,  F-91297 Arpajon, France}
\author{J. Terasaki}
\affiliation{Center for Computational Sciences, University of Tsukuba, Tsukuba, 305-8577, Japan}
\author{J. Engel}
\affiliation{Department of Physics and Astronomy, University of North Carolina, Chapel Hill, North Carolina 27599-3255}

\begin{abstract}
A microscopic calculation of reaction cross sections for nucleon-nucleus scattering has been performed by explicitly coupling the elastic channel  to all particle-hole excitations in the target and  one-nucleon pickup channels. 
The particle-hole states may be regarded as {\em doorway states} through which the flux flows to more complicated configurations, and 
subsequently to long-lived compound nucleus resonances. 
Target excitations for $^{40,48}$Ca, $^{58}$Ni, $^{90}$Zr and $^{144}$Sm were described in a random-phase framework using a Skyrme
functional.  
Reaction cross sections obtained agree very well with experimental data and predictions of a state-of-the-art fitted optical potential. 
Couplings between inelastic states were found to be negligible, 
while the pickup channels contribute significantly.
The effect of resonances from higher-order channels was assessed. Elastic angular distributions were also calculated within the same method, achieving good agreement with experimental data.
For the first time observed absorptions are completely accounted for by explicit channel coupling, 
for incident energies between 10 and 70 MeV, with consistent angular distribution results.
\end{abstract}
\date{\today}

\maketitle

\section{Introduction}

Achieving a quantitative and predictive description of the structure of and reactions with nuclei across the isotopic chart is an important and challenging goal of nuclear physics. Accurate knowledge of reaction rates, in particular those related to reactions induced by a single nucleon, are important for nuclear energy applications and for understanding astrophysical phenomena \cite{FrontiersNuclSci2007}, such as the evolution of stars and the synthesis of the elements. Radiobiology and space science developments rely on the proper determination of reaction observables to provide accurate yields and spectra for radiation protection and risk estimates \cite{Townsend2002ASR30}. National security applications also make use of reaction and structure information to detect nuclear materials of interest.

Accurate prediction of quantities related to nuclear reactions is a complex problem as not only the desired outcome, i.e.\ exit channel, has to be considered, but also the interference and competition with all other possible outgoing channels. 
A successful account of elastic nucleon-nucleus scattering, for example, has to include the effects from the excitation of non-elastic 
degrees of freedom, such as collective and particle-hole (p-h) excitations, transfer reactions, etc. 
The picture that emerges is one in which flux is removed from the elastic channel by couplings to the non-elastic degrees of freedom. 
Formally, these non-elastic
effects can be accounted for by the projection-operator approach of Feshbach \cite{Feshbach1958ARNS8}. 
This approach reduces the complexity of the problem by introducing an effective optical model potential (OMP). This potential, often complex, can be defined \cite{Feshbach1958ARNS8,Brown1959RMP31} as the effective interaction in a 
single-channel calculation that contains
the effects of all the other processes that occur during collisions between nuclei.
%
%
OMPs play a very important role in the description of nuclear reactions. They are extensively used to describe the interactions of
projectile and target in the entrance channel, and the interaction of ejectile and residual nuclei after the reaction; they are crucial
ingredients in direct-reaction analyses (e.g. elastic and inelastic scattering, transfer reactions, etc.) and provide transmission coefficients for statistical (Hauser-Feshbach) calculations. 

Most widely used are phenomenological OMPs fitted to reproduce experimental data sets. 
They have been extremely successful for many applications involving nuclei in the range of the fits \cite{Satchler1979PR55}. 
At the same time, such adjustable potentials make strong assumptions about locality and momentum dependence that are probably not justified. In addition, present nuclear theory applications require increasingly accurate predictions, specially for isotopes off stability. However, for nuclei lying outside the range of the fits, such as unstable nuclei produced at rare-isotope facilities, in the $r$-process, and in advanced reactor applications, this can lead to unquantifiable uncertainties. The reaction mechanisms of systems involving weakly-bound nuclei are known to be different from the strongly-bound ones \cite{Zerva2010PRC82,Nobre2007PRC76}. Studies have shown that the behaviour of phenomenological OMPs for stable systems usually cannot be directly extended to systems involving weakly-bound nuclei \cite{Deshmukh2011PRC83}.
To achieve a better understanding of nuclear reactions and structure, it is thus important to calculate OMPs by more fundamental, general, and first-principle methods.

According to microscopic reaction theory, an OMP is comprised of two components. 
The first is a real bare potential, corresponding to the diagonal potential within the elastic channel, which is generally obtained by folding the nucleon distributions of  both nuclei with a nucleon-nucleon effective interaction. 
The second is a complex dynamic polarization potential which arises from couplings to inelastic states. 
The resulting optical potential is complex, composed of a real part (usually slightly different from the bare potential) and an imaginary component. The latter gives rise to absorption of flux from the elastic channel to the other reaction channels, and is hence directly connected with observed reaction cross-sections.

Several attempts have been made to generate OMPs from microscopic approaches. Some have used the  nuclear matter approach \cite{Jeukenne1977PRC16}, where the calculation is first performed in nuclear matter and a potential is then obtained for finite nuclei by an appropriate local density approximation. This approach provides accurate results
at nucleon energies $\gtrsim$ 50 MeV \cite{Barbieri2005PRC72}. Recently, new methods based on self-energy theory have  been implemented \cite{Amos2000ANP25}, and new calculations, which
combine a nuclear matter approach and  Hartree-Fock-Bogoliubov (HFB) mean field structure model, provide encouraging results for neutron scattering below 15 MeV \cite{Pilipenko2010PRC81,Kuprikov2006PAN69}. Earlier attempts used the nuclear structure approach, which is more suitable at energies below ~50 MeV \cite{Bouyssy1981NPA371}, and calculated second-order diagrams using particle-hole propagators in the random-phase approximation (RPA) \cite{VinhMau1976NPA257,Bernard1979NPA327,Osterfeld1981PRC23,Bouyssy1981NPA371}. However, these were not able to fully explain observed absorption: e.g., in Ref.\ \cite{Bernard1979NPA327}, the couplings could account only for $\approx$ 44\% of the nucleon-nucleus absorption and, in Ref.\ \cite{Osterfeld1981PRC23}, only for $\approx$ 71\% including charge exchange.

The construction of OMPs from microscopic methods that use mapping of effective interactions to nucleon-nucleon \emph{g} matrices, which are solutions of nuclear matter equations, have proven to provide good agreement with reaction cross-section data \cite{Deb2001PRL86,Amos2002PRC65}. However, due to the increased number of discrete states of heavier nuclei ($A \gtrsim 12$), this approach does not accommodate their specific structure and processes. 
A method that implements successive spectator expansions of the optical potential, where the projectile is considered to interact at first order with a single nucleon of the target, has also proven to be a successful tool to obtain nucleon-nucleus cross sections \cite{Chinn1995PRC52}, although analyses were made only for scattering energies in the range $\sim$65 to 400 MeV. 
In addition, the method described in Ref. \cite{Chinn1995PRC52} treats the propagator modification through a nuclear mean field potential taken from structure calculations. Although this is a valid approach, it may not be completely satisfactory when aiming to keep full consistency with the theory of multiple scattering. A more consistent alternative, although not intractable, would be more difficult to implement. 

RPA-based microscopic methods have also been applied to heavy-ion reactions achieving good description of the double giant dipole resonance \cite{Lanza1997NPA613,Lanza1998NPA636,Andres2001PRC65}. In such works, one- and two-phonon states were populated and anharmonicities and nonlinear terms were treated. OMPs were then obtained through a semiclassical approach, by integrating the excitation probability over all impact parameters.

The latest advancements in the description of the structure of the nuclei from \emph{ab initio} methods \cite{UNEDF} allow for the development of more fundamental reaction models based directly on structure results. Such microscopic approach leads to the calculation of reaction observables that are consistent with the structure inputs adopted. In addition, OMPs based on microscopic approaches are much more reliable than phenomenological ones when extrapolated to describe processes involving unstable nuclei or previously unquantified transitions.
Among the different microscopic structure models, methods based on energy-density functionals (EDF) emerge as the only tractable theoretical tool that can be applied to all the nuclides with $A \gtrsim 40$ \cite{UNEDF}. 

In this paper we report on recent progress made towards achieving a complete microscopic calculation of the reaction cross sections for both neutron and proton induced reactions on a variety of medium-mass targets. We make extensive use of recently developed fundamental structure models based on energy-density-functional theory, such as RPA and quasi-particle RPA (QRPA). We calculate sets of excited states and the corresponding transition densities and potentials from the ground state (g.s.) for different nuclei across the periodic table. We then incorporate this information about the transitions in  coupled-reaction channels calculations of nucleon-nucleus reactions, coupling to all relevant channels necessary to consistently obtain accurate cross sections. First results of this approach were reported in Ref.~\cite{Nobre2010PRL105}, while here we detail and extend that work. 

The paper is structured as follows: in Section \ref{Sec:Structure} we present details of the structure models used and explain how they connect with reaction calculations; Section \ref{Sec:CRC} explains the procedure adopted for the coupled reaction channel calculations; in Section \ref{Sec:Results}  our main results are shown and discussed; and Section \ref{Sec:Conclusion} presents our final conclusions.

\section{Structure Models}
\label{Sec:Structure}

A HFB calculation gives  the particle and hole levels of a given nucleus and fixes the p-h basis states for generating excited states within the framework of (Q)RPA,  thus accounting for long range correlations caused by the residual interactions within the target.
To obtain the initially  
occupied proton and neutron levels in a nucleus, we use the Skryme energy-density functional SLy4 \cite[Table 1]{Chabanat1998NPA635}, a 
parametrization designed to  describe systems with arbitrary neutron excess, from stable to neutron matter, by improving isotopic properties, which overcomes deficiencies of other interactions away from the stability line. Although we used only the SLy4 parametrization in our work, the method is general enough to use any Skyrme force or any other functional.

\subsection{Ingredients of the Calculation}

In this subsection, we show the equation of the single-particle wave function and building block of the RPA excited state.
To describe the Hartree-Fock basis formed by particles with spin $s=\frac{1}{2}$, the following state vectors can be defined:
\be
| nl\tfrac{1}{2}jmt \rangle = \frac{\phi_{nljt}(r)}{r} \, i^l \, [Y_{l}(\mathbf{\hat{r}}) \times \chi_\frac{1}{2}]_{jm} \; \chi_{\frac{1}{2} t}^{iso},
\label{eq_rpa_expr_010}
\ee
with radial wave functions $\phi_{nljt}(r)$ expressed in a coordinate representation, and $\chi_\frac{1}{2}$ and $\chi_{\frac{1}{2} t}^{iso}$ are spin and isospin components, respectively.  We associate the creation and annihilation operators $a_{nlsjmt}^\dagger$ and $a_{nlsjmt}$ with these state vectors.  It is convenient to define a modified annihilation operator, $\tilde{a}_{nlsjmt} = \signfac{j-m} \, a_{nlsjmt}$, which is a spherical tensor of rank $j$ and projection $-m$.

We now define particle ($p$) and hole ($h$) states corresponding to orbitals above or below the Fermi surface, schematically indicated by $p>F$ and $h<F$, respectively.  The symbol $p$ represents all quantum numbers except the magnetic projection; \idest $p \equiv \{n_p l_p \frac{1}{2} j_p t_p\}$.  The same definition applies to the hole states, replacing $p$ by $h$.

With the above definitions we define an operator that creates a particle-hole pair coupled to angular momentum $J$ and projection $M$,
\be
\mathcal{A}_{JM}^\dagger(p,\tilde{h}) = \sum_{m_p m_h} (j_p m_p j_h -m_h | J M) \; a_{p m_p}^\dagger \tilde{a}_{h m_h}.
\label{eq_rpa_expr_030}
\ee
The Hermitian conjugate of $\mathcal{A}_{JM}^\dagger(p,\tilde{h})$, 
\be
\mathcal{A}_{J\bar{M}}(p,\tilde{h}) = \signfac{J-M} \; \mathcal{A}_{J-M}(p,\tilde{h}),
\label{eq_rpa_expr_040}
\ee
destroys a particle-hole pair, where $\bar{M}=-M$.
Both $\mathcal{A}_{JM}^\dagger(p,\tilde{h})$ and $\mathcal{A}_{J\bar{M}}(p,\tilde{h})$ are spherical tensors of rank $J$ and projection $M$.

The QRPA formalism extends this HFB approach by including quasiparticle excitations in the following way:
1) the single-particle wave function is extended to a two-component representation \cite{Dobaczewski1984NPA422}; 
2) the particle-creation-hole-annhilation and particle-annihilation-hole-creation are extended to two-quasiparticle creation and annihilation \cite{RingManyBody,Terasaki2005PRC71}. 

The HFB calculations were performed using a slightly modified version of the \textsc{hfbrad} \cite {Dobaczewski1984NPA422} code called \textsc{hfbmario} (version 6.2) \cite{StoitsovPrivate}.

\subsection{RPA States}

We define a boson operator $\Theta_{NJM}^\dagger$ that creates the RPA state $|NJM\rangle$, where $N$ is the principal quantum number, when applied to the correlated ground state $|0\rangle$, which is the vacuum for the RPA excitations.  We also define the time-reversed destruction operator
\be
\Theta_{NJ\bar{M}} = \signfac{J-M} \; \Theta_{NJ-M}.
\ee
The particle-hole operators relate to the boson operators through
\be
\begin{split}
\label{eq_rpa_expr_120}
\Theta_{NJM}^\dagger  
=  \sum_{p>F,\; h<F} X_{ph}^{NJ} \; \mathcal{A}_{JM}^\dagger(p,\tilde{h}) - Y_{ph}^{NJ} \; \mathcal{A}_{J\bar{M}}(p,\tilde{h}) ,
\end{split}
\ee
\be
\begin{split}
\label{eq_rpa_expr_130}
\Theta_{NJ\bar{M}}  =
  \sum_{p>F,\; h<F} X_{ph}^{NJ} \; \mathcal{A}_{J\bar{M}}(p,\tilde{h}) - Y_{ph}^{NJ} \; \mathcal{A}_{JM}^\dagger(p,\tilde{h})  ,
\end{split}
\ee
where $X_{ph}^{NJ}$ and $Y_{ph}^{NJ}$ are the components of the linear combination of particle-hole excitations used to construct the RPA states \cite{RingManyBody,Thouless1961NP22}. 
In the present treatment the $X$ and $Y$ coefficients are real and obey the normalization and orthogonality relations
\be
\sum_{p>F,\; h<F} X_{ph}^{NJ} X_{ph}^{N'J} - Y_{ph}^{NJ} Y_{ph}^{N'J} = \delta_{NN'}
\label{eq_rpa_expr_100}
\ee
and
\be
\sum_{p>F,\; h<F} X_{ph}^{N'J} Y_{ph}^{NJ} - X_{ph}^{NJ} Y_{ph}^{N'J} = 0.
\label{eq_rpa_expr_110}
\ee
Such $X$ and $Y$ amplitudes were obtained by solving the RPA equations \cite{Dupuis2006PRC73} with the particle-hole interaction derived from the second derivative of the energy density functional \cite{Dupuis2008PLB665}.

\subsection{RPA Transition Densities}

Transitions of the many-nucleon system from an initial RPA state $\vert i \rangle=\vert \alpha_iI_iM_i\rangle$, with angular momentum $I_i$ and projection $M_i$ to a final RPA state $\vert f \rangle=\vert \alpha_fI_fM_f\rangle$, where  $\alpha_{i}$ and $\alpha_{f}$ are additional quantum numbers required to characterize the states, can be expressed with the help of transition densities \cite{DupuisPhDthesis}:
\begin{equation}
\begin{split}
& \rho_{S \nu}^{T q, ~ fi}(\mathbf{r}_t)  \\
&= \langle \alpha_f I_f M_f | \sum_n \delta(\mathbf{r}_t{-}\mathbf{r}_n) \, \mathcal{S}_{S \nu}^n \mathcal{T}_{T q}^n | \alpha_i I_i M_i \rangle \\
& \equiv    4 \pi \!\! \sum_{L\mu JM} \!\! (L \mu S \nu | JM) (I_i M_i J M | I_f M_f)  \rho_{LSJ}^{T q, ~ fi}(r)  Y_{L\mu}^*(\mathbf{\hat{r}}_t) ,
\end{split}
\end{equation}
where the sum over index $n$ represents summation over all occupied orbitals (for protons or neutrons), $\mathbf{r}_t$ is the position at which the transition density is calculated, and $\mathbf{r}_n$ is the spatial-coordinate operator of the $n$'th particle in the Hilbert space containing the states. The $\mathcal{S}_{S \nu}^n$ and $\mathcal{T}_{T q}^n$ are the spin and isospin transition operators respectively; $T=0$ corresponds to the isoscalar part of the interaction and $T=1$ to the isovector one. For $S=0$ or~$1$,
\be
\mathcal{S}_{00}= 1 \mbox{~~~~ and ~~~~} \mathcal{S}_{1 \nu}= \sigma_{1 \nu}, 
\label{eq_eint_cent_020}
\ee
together with similar quantities $\mathcal{T}_{T q}$ for isospin,
\be
\mathcal{T}_{00}= 1 \mbox{~~~~ and ~~~~} \mathcal{T}_{1 q}= \tau_{1 q} , 
\label{eq_eint_cent_030}
\ee
where  $\sigma_{1 \nu}$ and $\tau_{1 q}$ are the spherical components of the vector of Pauli matrices. 
It is convenient to introduce a multipole expansion for the transition densities. The coordinate-space radial multipole transition density  is
\begin{equation}
\begin{split}
&\rho_{LSJ}^{T q, ~ fi}(r) = \frac{1}{\sqrt{2I_f{+}1}} \,\\
&\times ( \alpha_f I_f || \sum_n \, \frac{1}{4 \pi} \, \frac{\delta(r-r_n)}{r \, r_n} \;  
[Y_L(\mathbf{\hat{r}}_n) \times \mathcal{S}_S^n]_J \; \mathcal{T}_{T q}^n || \alpha_i I_i ).
\label{eq_trden_intro_120}
\end{split}
\end{equation}
Using second-quantization techniques, we can write Eq. (\ref{eq_trden_intro_120}) as
\begin{equation}
\begin{split}
\label{eq_trden_intro_210}
\rho_{LSJ}^{T q}(r) & = \sqrt{\frac{2I_i+1}{2I_f+1}} \sum_{\stackrel{\scriptstyle{\alpha_1 j_1 t_1}}{\scriptstyle{\alpha_2 j_2 t_2}}} Z_{t_2 t_1}^J(\alpha_2 j_2, \alpha_1 j_1) \\
&   \times ( \alpha_2 j_2 t_2 || f_L(r) [Y_L(\mathbf{\hat{r}}) \times \mathcal{S}_S ]_J || \alpha_1 j_1 t_1 ) \; \mathcal{T}_{Tq}^{t_2 t_1}, 
\end{split}
\end{equation}
where the isospin matrix element is $\mathcal{T}_{Tq}^{t_2 t_1} = \langle \smallfrac{1}{2} t_2 | \mathcal{T}_{Tq} | \smallfrac{1}{2} t_1 \rangle$
. In this expression the density is a sum over reduced matrix elements between single-particle states weighted by spectroscopic amplitudes $Z_{t_2 t_1}^J(\alpha_2 j_2, \alpha_1 j_1)$, the calculation of which will be described below.

We can calculate the corresponding momentum-space density  by the Fourier Bessel transform, as described in Appendix~\ref{FourBess}, obtaining
\begin{equation}
\begin{split}
&\rho_{LSJ}^{T q, ~ fi}(q) = \frac{1}{\sqrt{2I_f{+}1}} \\
&\times ( \alpha_f I_f || \sum_n \, j_L(qr_n) \;
 [Y_L(\mathbf{\hat{r}}_n) \times \mathcal{S}_S^n]_J \; \mathcal{T}_{T q}^n || \alpha_i I_i ).
\label{eq_trden_intro_180}
\end{split}
\end{equation}

Transition densities can be classified according to the specific states that they connect.
For a spin-0 even-even spherical nucleus, we can divide the problem of finding 
such 
densities into three parts, according to the number of RPA phonons in the initial and final states: \emph{1. No phonons in either initial or final state} (this is needed for elastic scattering from the ground state, described in Sec.~\ref{gs-gs}); \emph{2. No phonon in the initial state, one phonon in the final state} (this is needed for inelastic scattering from the ground state to an excited state, described in Sec.~\ref{gs-exc}); \emph{3. One phonon in each of the initial and final states} (this corresponds to inelastic scattering between two excited states, and also to elastic scattering from an excited state. It is described in Sec.~\ref{exc-exc}).

In each case the main work is the calculation of the spectroscopic amplitudes (or $Z$ coefficients).  Using these we construct the radial densities needed for folding-model calculations of transition and diagonal potentials.  Of course, the same coefficients can be used to calculate any other one-body operator connecting the initial and final states.

We use the symbol $p$ to represent the set of quantum numbers $\{n_p l_p \frac{1}{2} j_p t_p\}$ for particle states, along with an equivalent definition for hole states.  When the particle has a subscript ({\it e.g.}\ $p_1$), rather than use the cumbersome $p_1 \equiv \{n_{p_1} l_{p_1} \frac{1}{2} j_{p_1} t_{p_1}\}$ we use the simpler notation $\{n_1 l_1 \frac{1}{2} j_1 t_1\}$ unless it leads to ambiguity; a corresponding notation is used for hole states. The superscripts $n$ and $p$ of the densities below indicate that neutrons and protons orbital transitions have to be calculated independently, so they can be later combined to form the transition potentials.

\subsubsection{Elastic from ground state}
\label{gs-gs}

The transition density that describes elastic scattering from the 0$^{+}$ RPA ground state is given by:
\be
\begin{split}
\label{eq_rpa_ggdiag_060}
&\rho_{000}^{n(p),g.s.}(r)  =  \left( \frac{1}{4\pi} \right)^\frac{3}{2}  \\
&\times \left[  \ \sum_{p_1 > F, \ p_2 > F}\mbox{$\!\!\!\!\!\!\!\!^{'}$} \ \ \sqrt{2j_2+1} \right.  
 \left. \ Z_{tt,elas}^{g.s.}(p_2, p_1) \ \frac{\phi_{p_2}(r)}{r} \; \frac{\phi_{p_1}(r)}{r} \right.  \\
&\!\!\!+\!\! \left. \sum_{h_1 < F, \ h_2 < F}\mbox{$\!\!\!\!\!\!\!\!^{'}$} \ \ \sqrt{2j_2+1} \right.  
 \left.  \ Z_{tt,elas}^{g.s.}(h_2, h_1) \ \frac{\phi_{h_2}(r)}{r} \; \frac{\phi_{h_1}(r)}{r} \right], 
\end{split}
\ee
where the prime on the summation sign is a indication that only neutron orbitals are to be used for the neutron density, and only proton orbitals for the proton density. Although in the elastic scattering the occupied orbitals do not correspond to holes, we still associate such states with the indices $h_1$ and $h_2$ in our notation to maintain consistency.

The ground-state density is easily evaluated from the above in the Hartree-Fock approximation.  The $Z$ coefficients for the particle states vanish, and those for the occupied states are simply $Z_{tt,elas}^{g.s.}(h_2, h_1) = \delta_{h_2 h_1} \sqrt{2j_2+1}$. Using this in Eq.~(\ref{eq_rpa_ggdiag_060}) we obtain that the density is
\be
\rho_{000}^{n(p),g.s.}(r) \ = \ \left( \frac{1}{4\pi} \right)^\frac{3}{2} \sum_{h<F}\mbox{$^{'}$} (2j_h+1) \left[ \frac{\phi_{h}(r)}{r} \right]^2.
\label{eq_rpa_ggdiag_080}
\ee

\subsubsection{From ground state to excited states}
\label{gs-exc}

After calculating the corresponding spectroscopic amplitudes ($Z$ coefficients) we have that the transition density needed for scattering from ground to an excited state described by a single RPA boson is
\be
\begin{split}
\label{eq_rpa_eg_120}
& \rho_{LSJ}^{n(p)}(r) \  =  \ \frac{1}{4\pi} \frac{1}{\sqrt{2J+1}} \\
 & \times \sum_{\stackrel{\scriptstyle{p>F}}{\scriptstyle{h<F}}}  \mbox{$^{'}$} \left[ X_{ph}^{NJ} + \signfac{S} \; Y_{ph}^{NJ} \right]  
 i^{l_h-l_p} \; \\
 & \times ( l_p \smallfrac{1}{2} j_p || [Y_L(\mathbf{\hat{r}}_t) \times \mathcal{S}_S ]_J || l_h \smallfrac{1}{2} j_h )  
 \; \frac{\phi_{p}(r)}{r} \; \frac{\phi_{h}(r)}{r}
\end{split}
\ee
where the initial (ground) state has spin 0.  

The single-particle matrix elements are:
\begin{multline}
\lefteqn{(l_2 \smallfrac{1}{2} j_2 || [Y_L \times \mathcal{S}_S ]_J || l_1 \smallfrac{1}{2} j_1 ) = \frac{1}{\sqrt{2 \pi}} \signfac{l_2} } \\ 
\times \, \hat{l_1} \hat{l_2} \hat{j_1} \hat{j_2} \hat{L} \hat{S} \hat{J} \threej{l_2}{l_1}{L}{0}{0}{0} \ninej{l_2}{l_1}{L}{\frac{1}{2}}{\frac{1}{2}}{S}{j_2}{j_1}{J}. 
\label{eq_twoquant_spmat_090}
\end{multline}
The following symmetry relation results from interchanging initial- and final-state quantum numbers:
\begin{multline}
\lefteqn{(l_1 \smallfrac{1}{2} j_1 || [Y_L \times \mathcal{S}_S ]_J || l_2 \smallfrac{1}{2} j_2 ) = }  \\
 \signfac{L+S+J+j_2-j_1} (l_2 \smallfrac{1}{2} j_2 || [Y_L \times \mathcal{S}_S ]_J || l_1 \smallfrac{1}{2} j_1 ).
\label{eq_twoquant_spmat_09}
\end{multline}

\subsubsection{Between excited states}
\label{exc-exc}

For scattering involving two excited states, the spectroscopic amplitudes may be divided into two components, identified with the superscripts $(1)$ and $(2)$. The parts of the spectroscopic amplitudes with superscript $(2)$  are present only if the initial and final states are the same and $J=0$.
After a great deal of angular momentum algebra, we find the parts of the spectroscopic amplitudes:
\be
\begin{split}
\label{eq_rpa_ee_170}
 Z_{tt}^{J}(p_2,p_1)^{(1)}& = \signfac{J_i - J} \, \sqrt{2J_f+1} \ \sum_h \signfac{j_2 + j_h}  \\
& \times \left[ X_{p_2 h}^{N_f J_f} \; X_{p_1 h}^{N_i J_i} \sixj{J_f}{J}{J_i}{j_1}{j_h}{j_2}  \right.   \\ 
&+\left. \signfac{J} \; Y_{p_1 h}^{N_f J_f} \; Y_{p_2 h}^{N_i J_i} \sixj{J_f}{J}{J_i}{j_2}{j_h}{j_1} \right] \\[.25in]
\end{split}
\ee
\be
\begin{split}
\label{eq_rpa_ee_180}
 Z_{tt}^{J}(h_2,h_1)^{(1)} &= - \signfac{J_f} \, \sqrt{2J_f+1} \ \sum_p \signfac{j_1 + j_p}  \\
& \times \left[ X_{p h_1}^{N_f J_f} \; X_{p h_2}^{N_i J_i} \sixj{J_f}{J}{J_i}{j_2}{j_p}{j_1}  \right. \\
& + \left. \signfac{J} \; Y_{p h_2}^{N_f J_f} \; Y_{p h_1}^{N_i J_i} \sixj{J_f}{J}{J_i}{j_1}{j_p}{j_2} \right] 
\end{split}
\ee

and

\be
\begin{split}
\label{eq_rpa_ee_210}
Z_{tt}^{J}(p_2,p_1)^{(2)} & =  \delta_{N_f J_f M_f, N_i J_i M_i} \\
&\times \ \delta_{JM, 00} \ Z_{tt,elas}^{g.s.}(p_2, p_1)
\end{split}
\ee
\be
\begin{split}
\label{eq_rpa_ee_220}
Z_{tt}^{J}(h_2,h_1)^{(2)} & =  \delta_{N_f J_f M_f, N_i J_i M_i}  \\
& \times \ \delta_{JM, 00} \ Z_{tt,elas}^{g.s.}(h_2, h_1)
\end{split}
\ee
where $N_{i}$ and $N_{f}$ are the principal quantum numbers of the initial and final states, respectively, and $J_{i}$ and $J_{f}$ are their corresponding angular momenta.

The corresponding transition density is:
\bea
\label{eq_rpa_ee_240}
& &  \rho_{LSJ}^{n(p)}(r) = \frac{1}{4\pi} \sqrt{\frac{2J_i+1}{2J_f+1}} \nonumber \\
& & \qquad \times \Bigg\{ \sum_{p_1 > F, \ p_2 > F}\mbox{$\!\!\!\!\!\!\!\!^{'}$} \ \ \left[ Z_{tt}^{J}(p_2,p_1)^{(1)} + Z_{tt}^{J}(p_2,p_1)^{(2)} \right] \nonumber \\
& & \times i^{l_1-l_2} \; ( l_2 \smallfrac{1}{2} j_2 || [Y_L(\mathbf{\hat{r}}_t) \times \mathcal{S}_S ]_J || l_1 \smallfrac{1}{2} j_1 ) \; \frac{\phi_{p_2}(r)}{r} \; \frac{\phi_{p_1}(r)}{r} \nonumber \\[0.1in]
& & \qquad \ \ + \sum_{h_1 < F, \ h_2 < F}\mbox{$\!\!\!\!\!\!\!\!^{'}$} \ \ \left[ Z_{tt}^{J}(h_2,h_1)^{(1)} + Z_{tt}^{J}(h_2,h_1)^{(2)} \right] \nonumber \\
& &  \times i^{l_1-l_2} \; ( l_2 \smallfrac{1}{2} j_2 || [Y_L(\mathbf{\hat{r}}_t) \times \mathcal{S}_S ]_J || l_1 \smallfrac{1}{2} j_1 ) \; \nonumber \\
&&  \times \frac{\phi_{h_2}(r)}{r} \;   \frac{\phi_{h_1}(r)}{r} \Bigg\}, 
\eea
where the reduced matrix elements are given by Eq.~(\ref{eq_twoquant_spmat_090}).  For transitions between different initial and final states, only the $Z$ coefficients with superscript $(1)$ appear.  When the initial and final states are the same, those with superscript $(2)$ also appear, but only for zero angular momentum transfer ($J=0$).  This case corresponds to elastic scattering and we see that the density is the same as for elastic scattering from the ground state, but with a correction term given by the coefficient with superscript $(1)$.  When $J$ is nonzero and the initial and final states are the same, only the coefficient with superscript $(1)$ appears and this can connect different magnetic substates when $L$ is even and overall angular momentum is conserved. This is a reorientation effect.

%

\subsection{Transition Potentials}
\label{TrPot}

The transition potential between any pair of levels of a given target nucleus is  obtained by a linear combination of the transition potentials for protons and neutrons of the target, which are calculated by folding an effective nucleon-nucleon interaction with the corresponding transition density. 
This linear combination depends on whether the projectile is a neutron or a proton. If $ v^{ST}(q)$ is the Fourier transform of the effective interaction and $\rho_{LSJ}^{Tq, ~ fi, (n)}(q) $ and $\rho_{LSJ}^{Tq, ~ fi, (p)}(q) $ are the transition densities in configuration space for protons an neutrons respectively, then the transition potential $U_{LSJ}^{Tq, ~ fi}(q)$ for a (n,n) reaction is 
\be
U_{LSJ}^{Tq, ~ fi}(q) = (v^{S0}+v^{S1})\rho_{LSJ}^{Tq, ~ fi, (n)}+(v^{S0}-v^{S1})\rho_{LSJ}^{Tq, ~ fi, (p)}.
\ee
For a (p,p) reaction, the transition potential is
\be
U_{LSJ}^{Tq, ~ fi}(q) =  (v^{S0}-v^{S1})\rho_{LSJ}^{Tq, ~ fi, (n)}+(v^{S0}+v^{S1})\rho_{LSJ}^{Tq, ~ fi, (p)}.
\ee

Our scattering effective nucleon-nucleon interaction is of Gaussian shape, with parameters matched to  the volume integral and r.m.s.\ radius of the M3Y interaction at 40 MeV; it  includes a knock-on exchange correction \cite{Love}:
\be
v^{T}(r)= V_{0}^{T}  e^{-\mu_{T}^{2}r^{2}}.
\label{Eq:GaussVr}
\ee
In the momentum space, the explicit form of the effective interaction is
\be
v^{T}(q)= V_{0}^{T} \frac{1}{\mu_{T}^{3}} \pi^{3/2} e^{-q^{2}/(2\mu_{T})^{2}}
\label{Eq:GaussVq}
\ee
where we used the values  $V_0^{0} = -24.1921$ MeV and $\mu_{0} = 0.7180$ fm$^{-1}$ for the isoscalar part of the interaction and $V_0^{1} = 11.3221$ MeV and $\mu_{1} = 0.7036$ fm$^{-1}$ for the isovector component
. We do not include any imaginary part in this effective interaction, as our aim is to include all nonelastic excitations explicitly in our model.
Using the Fourier transform $ v^{T}(q)$ of the effective interaction, the configuration-space transition potential
is
\be
U_{LSJ}^{Tq, ~ fi}(r) = \frac{2}{\pi} \int_0^\infty dq \; q^2 \; j_L(qr) \; v^{T}(q) \, \rho_{LSJ}^{Tq, ~ fi}(q) .
\label{eq_dwa_trpo_090}
\ee

For the reaction calculations coupling to QRPA states, we used the single-folded potential constructed by folding the interaction of Eq.~(\ref{Eq:GaussVq}) with the ground-state density from the HFB calculation as the bare potential in the elastic channel. For simplicity, this was also used as diagonal potential for all excited states as well. For RPA couplings we used Eqs.\ (\ref{eq_rpa_ggdiag_080}) and (\ref{eq_rpa_ee_170})-(\ref{eq_rpa_ee_240}) to obtain the diagonal potentials. Diagonal potentials describe the elastic scattering for the g.s. or an excited state, while the off-diagonal potentials provide the couplings between the different states of the nuclei.

\section{Coupled Reaction Channels Calculations}
\label{Sec:CRC}

According to the coupled reaction channels (CRC) formalism, to correctly account for the effects of the competition between the different processes occurring in a nuclear reaction, all possible couplings between channels (elastic, inelastic, transfer, etc.) should be explicitly considered. However, in practice the number of couplings has to be limited, taking into account the relevance of each channel. A criterion that may be adopted to define this limit is to consider the corresponding order of the correction to the elastic scattering. 
To explore the relative importance of the various contributions to the reaction cross section, we carried out a series of calculations including three different sets of couplings: A) only inelastic couplings from and to the ground state, described in Sec. \ref{Sec:Inelastic}; B) same couplings as in A) but also couplings between the excited states, described in Sec. \ref{Sec:BetweenExc}; and C) same as in A) but also couplings to transfer channels leading to the formation of a deuteron, described in Sec. \ref{Sec:Transfer}. Table \ref{Tab:Calculations} illustrates the different couplings that were explicitly included in the three sets of calculations (A, B, and C), and the corresponding order of the correction from elastic scattering. The contribution of the third-order processes included in calculation B) was found to be negligible, as shown in Sec. \ref{Sec:BetweenExc}. Hence, other third-order (i.e. couplings between inelastic and transfer channels) and fourth-order (deuteron break-up channels, multi-step transfer via continuum) couplings were neglected.

\begin{table}[h]
\begin{center}
\caption{Summary of the channels  that were explicitly coupled in the calculations A, B, and C. The corresponding order of the correction relative to elastic scattering is also shown.}
\label{Tab:Calculations}
\begin{tabular}{lcccr}
\hline
\hline
 Couplings present & A & B & C & Correction\\
\hline
Excited states with g.\ s.     & Yes & Yes & Yes & 2$^{nd}$ order \\
Between excited states     & No & Yes & No &      3$^{rd}$ order  \\
Transfer with deuteron channel & No & No & Yes & 2$^{nd}$ order  \\
\hline
\hline
\end{tabular}
\end{center}
\end{table}


Truncations in channel space are necessary in any practical calculation. 
In our calculations, we depend only on real components in the scattering potentials, as we aim to reduce the number of phenomenological parameters to a minimum. We therefore exclude all channels coupled to the elastic at second order and above, and use the doorway approximation so that the imaginary parts of the optical potentials in the first-order channels hardly affect  the scattering in the elastic channel. Any imaginary potentials introduced into our models are only for computational convenience. They allow us to temporarily replace couplings already studied, and to focus on additional effects.

\subsection{Inelastic Coupled Channels}
\label{Sec:Inelastic}

We initially performed coupled channels calculations for reactions involving protons and neutrons scattered by the nuclei $^{40}$Ca, $^{48}$Ca, $^{58}$Ni, $^{90}$Zr, and $^{144}$Sm, coupling the ground state to all levels with excitation energy ($E^{*}$) lying below some limit. Such excited states were obtained according to the QRPA model in a box of 15~fm.  The QRPA states above the particle emission threshold are used to approximate exact scattering waves. Recent studies  have shown that such wave functions contain large density distributions outside the nuclear radius \cite{Terasaki2007PRC76}. When used in reaction calculations they accurately represent the continuum \cite{Moro2007PRC75,Rodriguez-Gallardo2008PRC77,Lay2010PRC82} inside the maximum radius. Thus, processes containing one nucleon in the continuum (plus the inelastically scattered projectile) are included in our model. Several applications of this approach can be found in the literature for nuclear structure \cite{Hiyama2003PPNP51} and reactions \cite{Matsumoto2004PRC70} problems. Results for the nonelastic absorption for neutron-induced reaction on a $^{90}$Zr target, corresponding to the inelastic couplings from g.s., are shown in Figure~\ref{Fig:BetweenExcStates} (solid line).

\subsection{Couplings Between Inelastic Channels}
\label{Sec:BetweenExc}

\begin{figure}[h]
 \begin{center}
  \includegraphics[trim = 0mm 0mm -0mm 0mm, clip,scale=0.40,angle=0.0]{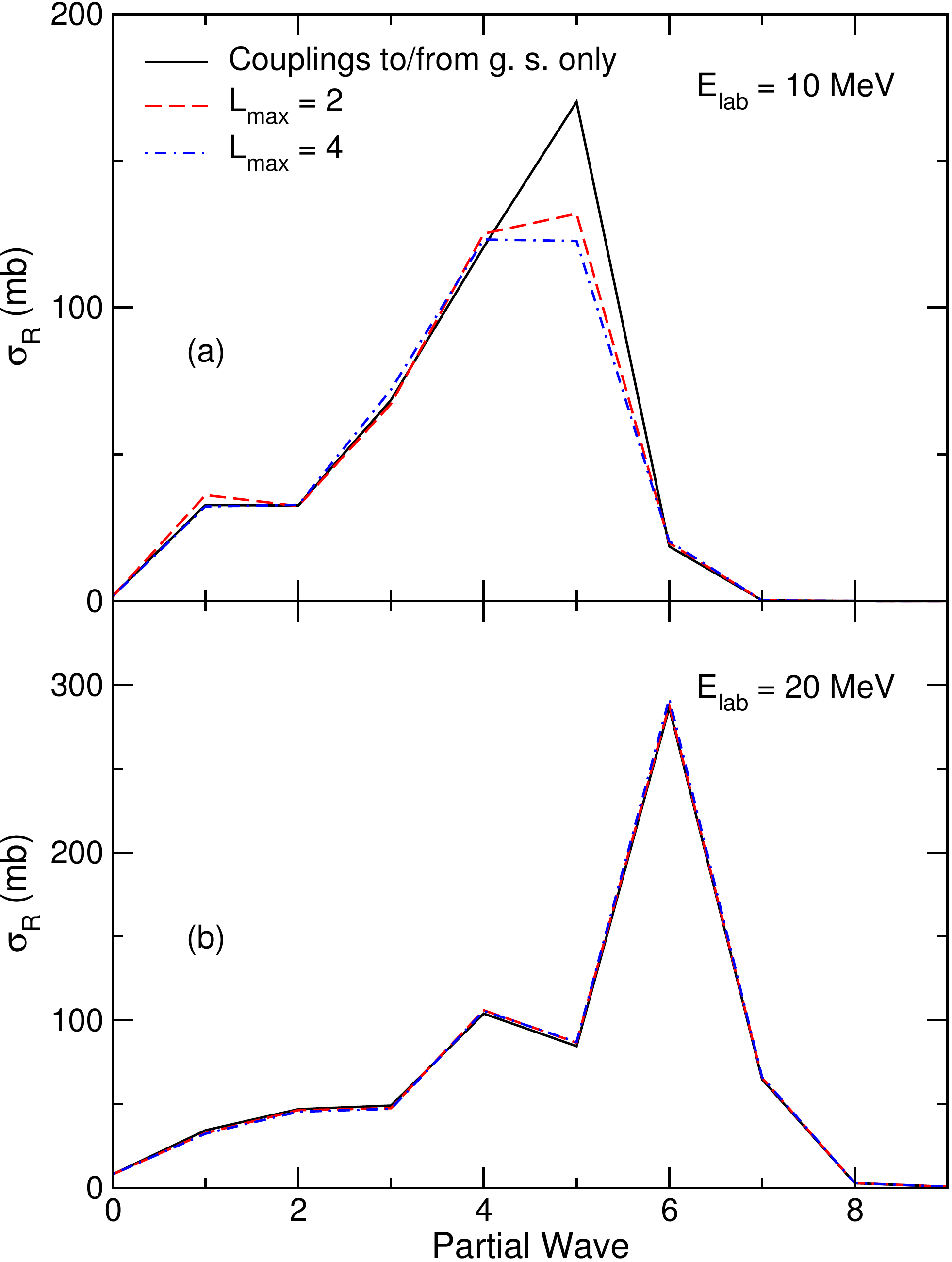} \\
 \end{center}
 \vspace{-4mm}
 \caption{(color online) Reaction cross-section as a function of the partial wave for the reaction n + $^{90}$Zr. Couplings to all RPA states below 20 MeV were included. The solid lines represent calculations without couplings between excited states. The dashed lines correspond to the  results when transitions between excited states are explicitly calculated, with maximum transferred angular momentum $L_{max}=2$. The dash-dotted lines correspond to the same calculation as the dashed ones but with $L_{max}=4$.}
 \label{Fig:BetweenExcStates}
\end{figure}

Couplings \emph{between} excited states, as predicted by the RPA model, were explicitly considered  with transition densities given by Eqs.\ (\ref{eq_rpa_ee_170})-(\ref{eq_rpa_ee_240}), for nucleons scattered by $^{90}$Zr. In Figure~\ref{Fig:BetweenExcStates} we show the reaction cross-section as a function of partial wave for the reaction n + $^{90}$Zr at scattering energies of 10 MeV (upper panel) and 20 MeV (lower panel), where all RPA states lying below 20 MeV were coupled. For each $E_{\mathrm{lab}}$ we compare the calculations considering only couplings to and from the ground state with calculations that also include couplings between excited states (with a maximum value, $L_{\mathrm{max}}$,  for the transferred angular momentum between them). It is observed that, although for $E_{\mathrm{lab}}=10$ MeV there are small but noticeable differences between calculations, for higher energies the curves are almost undistinguishable. 
Despite the fact that, above 10 MeV, the couplings between excited levels do change the cross-sections of the individual states (mostly small changes, although for some few channels the cross-sections may differ by a factor up to 20\%), the overall sum among all states remains unchanged. This supports the validity of the \emph{doorway approximation}, according to which  the total flux leaving the elastic channel to all possible first-order channels is independent of what happens afterwards:  a nucleon later might escape as a free nucleon, the flux might equilibrate to compound-nuclear resonances, etc. The fact that couplings between excited states were found to provide negligible contribution to the absorption summed over all states, for scattering energies above 10 MeV, allowed us to disregard them, and other higher-order multi-step processes, in the subsequent CC calculations.

\subsection{Coupling to Pick-up Channels}
\label{Sec:Transfer}

Pick-up channels  play an important role in nucleon-nucleus scattering \cite{Keeley2008PRC77,Mackintosh2007PRC76,Coulter1977NPA293}. 
We performed coupled reaction channels (CRC) calculations that included all open channels that describe the formation of a deuteron by picking up the appropriate nucleon from occupied levels in the target.
For transfers, we approximate the HFB target states by bound single-particle states in a Woods-Saxon potential,  with the radii fitted to reproduce the volume radii and Fermi energy obtained by the HFB calculations.
The real diffuseness and spin-orbit parameters were taken from Koning-Delaroche optical potentials \cite{Koning2003NPA713} 
at $E_{\rm lab}=0$, with spin-orbit radii adjusted by the same factor used to fit the volume part to HFB radii.
The values used  are shown on Table \ref{Tab:BindingPot}.

\begin{table}[h]
\begin{center}
\caption{Parameters of the binding potentials used for the coupled reaction channel (CRC) calculations}
\label{Tab:BindingPot}
\begin{tabular}{ccccc}
\hline
\hline
 Reaction&$a_{V} (fm)$&$V_{SO}$ (MeV)&$r_{SO} (fm)$&$a_{SO} (fm)$\\
\hline \vspace{-3mm} \\ 
$^{40}$Ca(n,d)     & 0.67 & 5.85 & 1.10 & 0.59  \\
$^{40}$Ca(p,d)     & 0.67 & 5.80 & 1.12 & 0.59  \\
$^{48}$Ca(n,d)     & 0.67 & 5.88 & 1.13 & 0.59  \\
$^{48}$Ca(p,d)     & 0.67 & 5.82 & 1.08 & 0.59  \\
$^{58}$Ni(n,d)      & 0.67 & 5.91 & 1.08 & 0.59  \\
$^{58}$Ni(p,d)      & 0.67 & 5.86 & 1.10 & 0.59  \\
$^{90}$Zr(n,d)      & 0.66 & 6.01 & 1.11 & 0.59  \\
$^{90}$Zr(p,d)      & 0.66 & 5.97 & 1.07 & 0.59  \\
$^{144}$Sm(n,d) & 0.66 & 6.19 & 1.10 & 0.59  \\
$^{144}$Sm(p,d) & 0.66 & 6.16 & 1.08 & 0.59  \\
\hline
\hline
\end{tabular}
\vspace{-2mm}
\end{center}
\end{table}

To overcome numerical limitations, we coupled explicitly only to the transfer channels, incorporating all inelastic effects in an inelastic optical potential obtained from coupling only to inelastic channels. The imaginary component of this inelastic optical potential corresponds to the Koning-Delaroche  \cite{Koning2003NPA713}  optical potential renormalized to account only for the inelastic absorption obtained in Calculation A) (described in Section \ref{Sec:Inelastic}).

CRC calculations require, in addition to the scattering potentials in the incoming channel, a scattering potential between the deuteron and the remaining target.
We adopted the Johnson-Soper \cite{Johnson1970PRC1} prescription as it includes the effects of deuteron breakup in adiabatic (sudden) approximation. 
In this prescription, the deuteron potential is the sum of the individual  neutron and proton potentials with the target. 
For the real parts we used the diagonal transition potentials  of the corresponding nucleon-nucleus reaction and,
for the imaginary parts, the sum of the  imaginary parts of the Koning-Delaroche  \cite{Koning2003NPA713}  optical potential for protons and neutrons. That is, fitted parameters are used in the imaginary part of the deuteron potential. A phenomenological deuteron potential, such as the one proposed by Daehnick \emph{et al.} \cite{Daehnick1980PRC21}, was not used since it  would introduce fitted parameters also in the real part of the deuteron potential. An iterative method of eliminating this need of phenomenological parameters is currently being investigated but we leave for future work to calculate the deuteron and nuclear potentials self-consistently.

\section{Results}
\label{Sec:Results}

To assess the success of our large-scale coupled-channels approach, we compare the calculated reaction cross section to that obtained by the Koning-Delaroche optical potential \cite{Koning2003NPA713}, which is one of the best nucleon-nucleus phenomenological optical potentials available, henceforth referred to as $\sigma_{\mathrm{R}}^{\mathrm{OM}}$.

\begin{figure}[h]
 \begin{center}
  \includegraphics[trim = 0mm 0mm 0mm 0mm, clip,scale=0.33,angle=0.0]{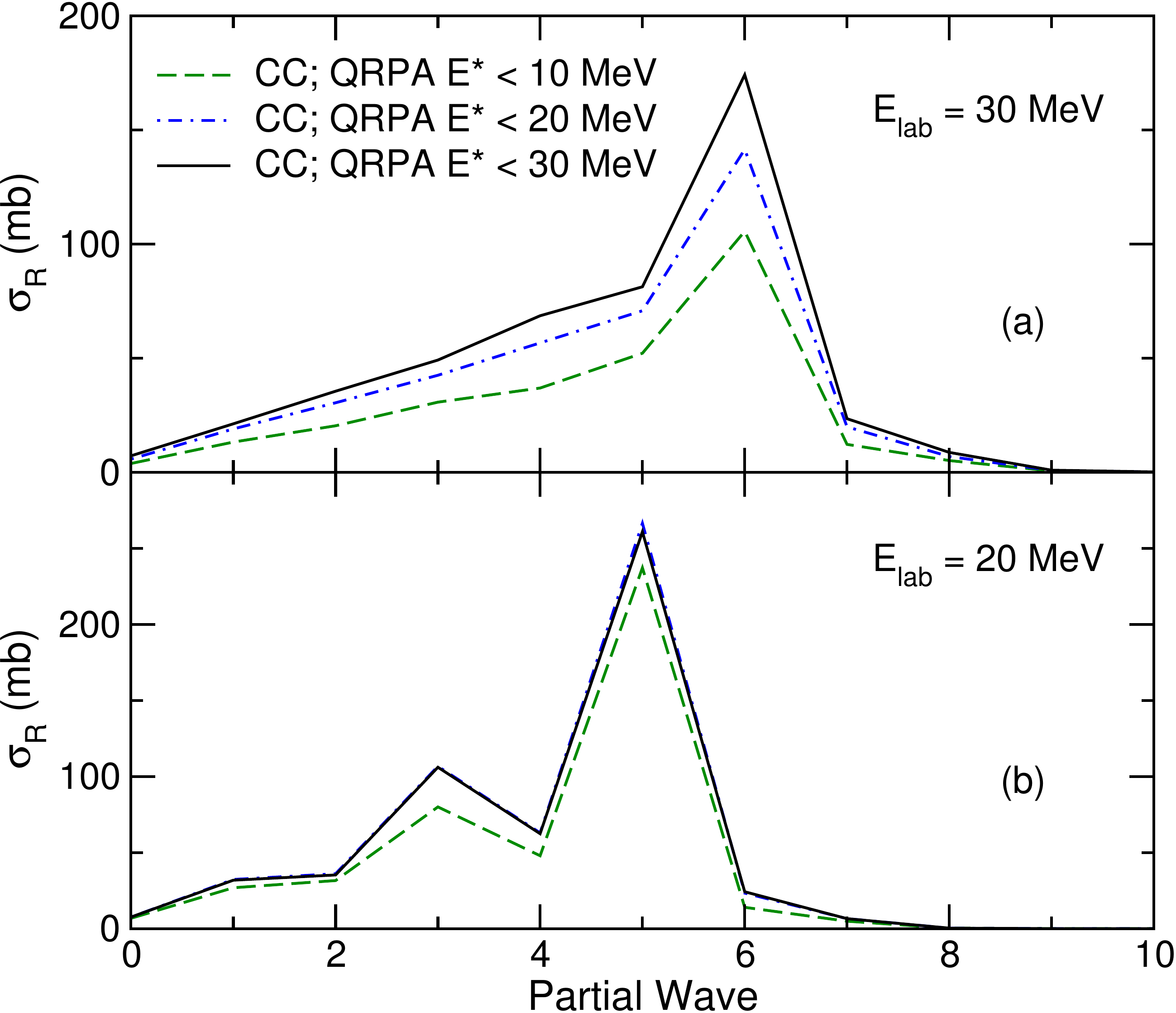} \\
 \end{center}
 \vspace{-4mm}
 \caption{(color online) Reaction cross-section as a function of the partial wave for the reaction n + $^{58}$Ni  at $E_{lab}$ = 30 MeV (a) and 20 MeV (b).}
 \label{Fig:nNi58SigRxJInel}
\end{figure}

We examined the convergence with respect to maximum excitation energy by performing CC calculations that explicitly couple to all QRPA excited states below the cutoff energies of 10, 20 and 30 MeV, for the incident energies $E_{\rm lab}$ = 10, 20, 30 and 40 MeV
. We found that convergence of the inelastic calculations requires coupling of  all excited levels below the scattering energy (i.e.\ all open channels). 
This behavior is illustrated  in Figure \ref{Fig:nNi58SigRxJInel}, for neutrons scattered by $^{58}$Ni. As it is seen in Figure \ref{Fig:nNi58SigRxJInel}, upper panel, coupling to more highly excited states results in larger reaction cross sections.
In principle, virtual couplings between the g.\ s. and closed channels can affect the reaction cross sections by changing the real, but not the imaginary, part of the OMP.  However, in the lower panel of Figure \ref{Fig:nNi58SigRxJInel} we observe that the couplings to states lying above the scattering energy do not noticeably contribute to the absorption.

\begin{figure}[h]
 \begin{center}
  \includegraphics[trim = 0mm 0mm 0mm 0mm, clip,scale=0.39,angle=-0.0]{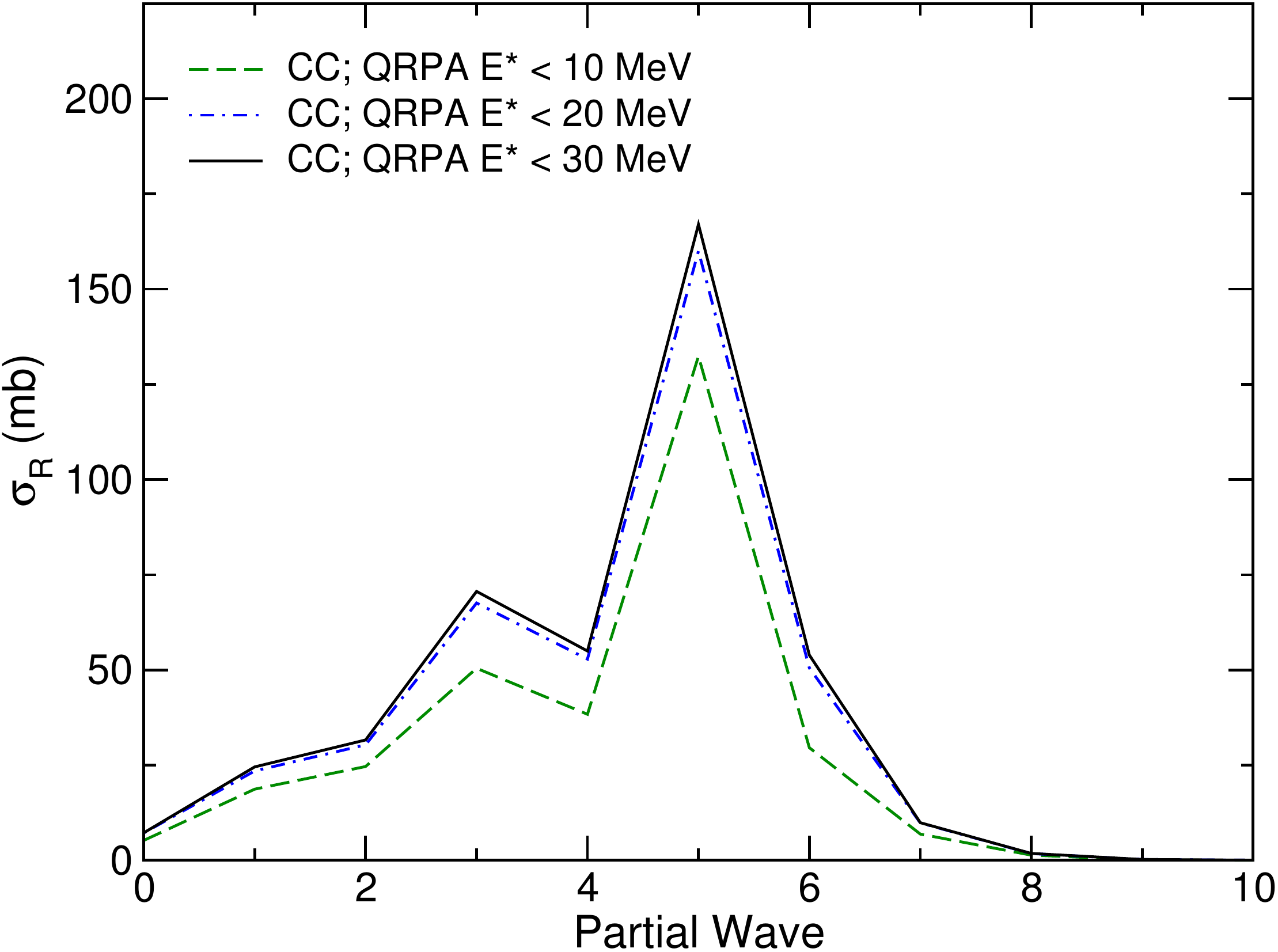} \\
 \end{center}
 \vspace{-4mm}
 \caption{(color online) Reaction cross-section as a function of the partial wave for the reaction p + $^{58}$Ni  at $E_{lab}$ = 30 MeV.}
 \label{Fig:pNi58SigRxJInel}
\end{figure}

\begin{figure}[h]
 \begin{center}
  \includegraphics[trim = 0mm 0mm 0mm 0mm, clip,scale=0.37,angle=-0.0]{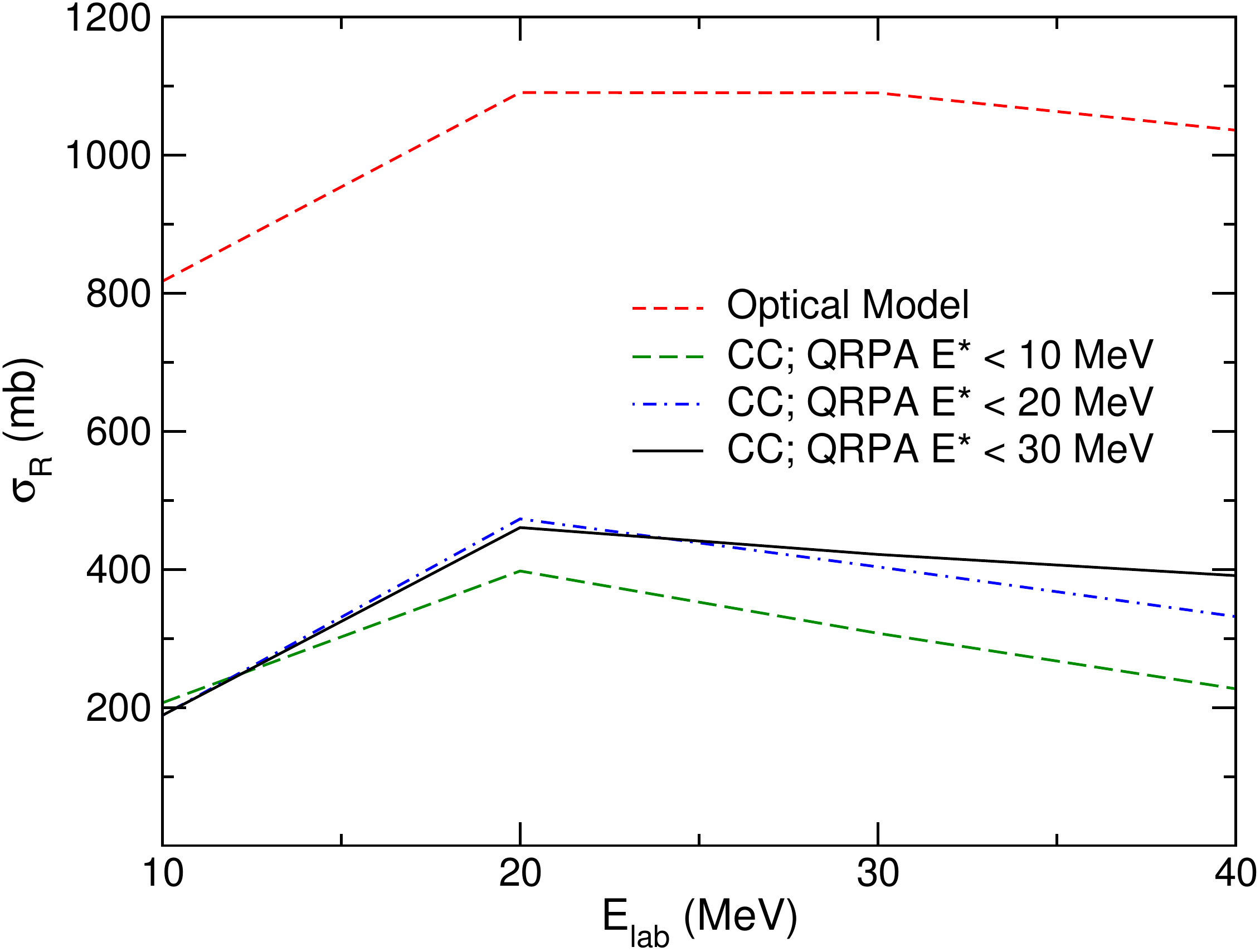} \\
 \end{center}
 \vspace{-4mm}
 \caption{(color online) Total reaction cross-section as a function of the incident energy for the reaction p + $^{58}$Ni, for the different inelastic calculations. The short-dashed line shows the results using the Koning-Delaroche \cite{Koning2003NPA713} optical potential. The lines serve as guides to the eye as calculations were performed only for $E_{\mathrm{lab}}$ = 10, 20, 30 and 40 MeV.}
 \label{Fig:pNi58SigRxElab}
\end{figure}

For reactions having protons  instead of neutrons as projectile, such inelastic convergence is reached with lower cutoffs, as is observed in Figure \ref{Fig:pNi58SigRxJInel} for p + $^{58}$Ni. This is due to the fact that a charged particle has to overcome the Coulomb barrier, which decreases the amount of energy available to excite target states. Thus, states close to $E_{\mathrm{lab}}$ that would correspond to open channels in the case of an incoming neutron become effectively closed for proton projectiles. This behavior is observed for each partial wave as well as at all energies, as is illustrated for p + $^{58}$Ni in Figures  \ref{Fig:pNi58SigRxJInel} and \ref{Fig:pNi58SigRxElab}, respectively. Figure \ref{Fig:pNi58SigRxJInel} shows the reaction cross section as a function of the partial wave for incident protons of 30 MeV. We observe that the difference between the results when coupling to states up to 20 and 30 MeV is much smaller than the one for  the similar system having neutrons as projectile, shown in Figure \ref{Fig:nNi58SigRxJInel} (upper panel). We also observe in Figure \ref{Fig:pNi58SigRxElab} that the effects of coupling to states lying between 20 and 30 MeV become noticeable only for $E_{\mathrm{lab}} \gtrsim$ 27 MeV, which is consistent with the Coulomb barrier height of $\approx$ 7.4 MeV, for this reaction.

\begin{figure}[t]
 \begin{center}
  \includegraphics[trim = 0mm 0mm 0mm 0mm, clip,scale=0.40,angle=-0.0]{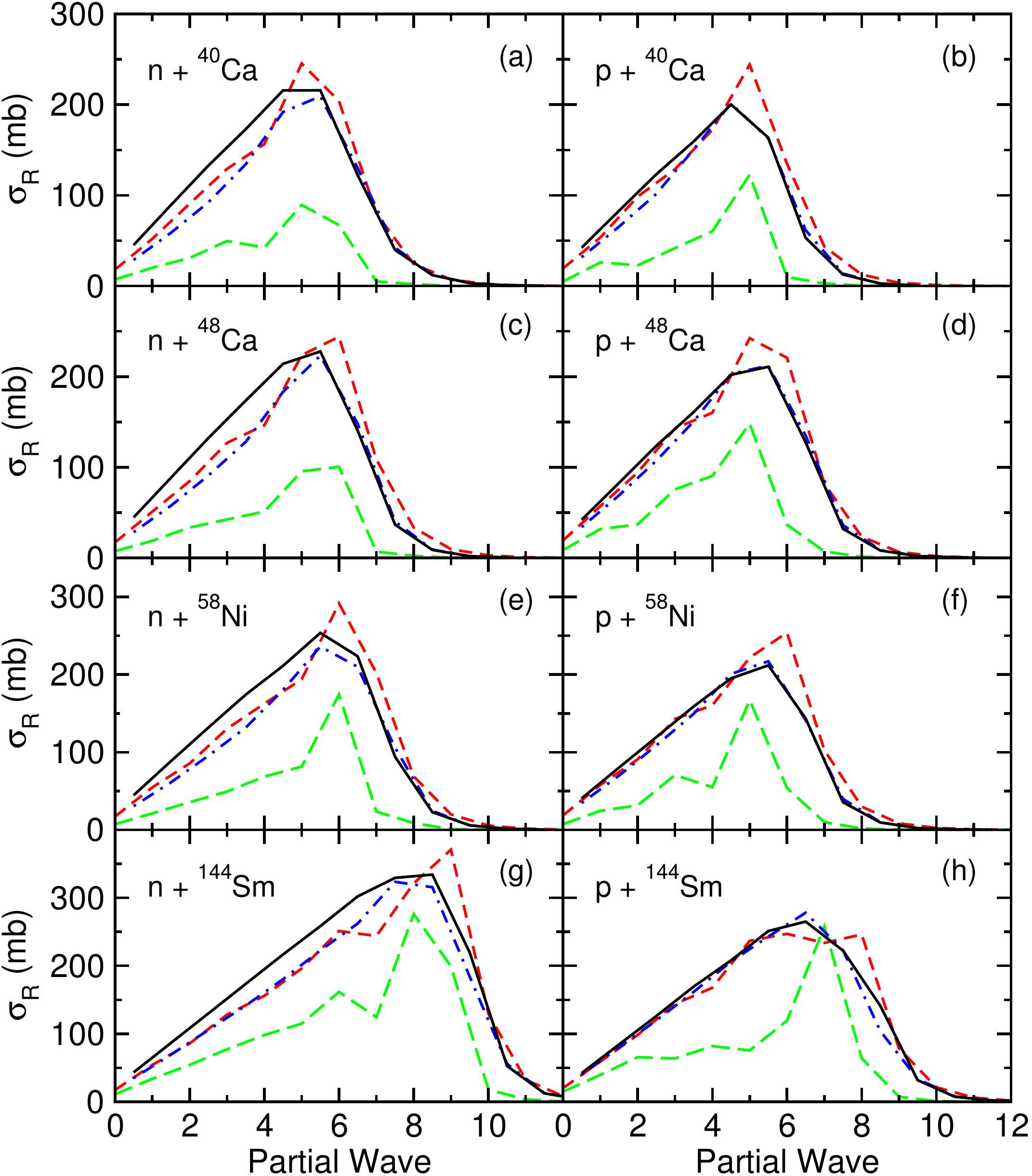} \\
 \end{center}
 \vspace{-4mm}
 \caption{(color online) Reaction cross-section as a function of the partial wave for the reactions n, p + $^{40,48}$Ca, $^{58}$Ni, $^{144}$Sm, for the different calculations at $E_{\mathrm{lab}}$ = 30 MeV. The results shown include couplings to the inelastic states lying below 30 MeV (dashed green lines), to the inelastic and transfer  channels (dash-dotted blue lines) and to the inelastic and transfer channels with non-orthogonality corrections (solid black lines). The Koning-Delaroche  \cite{Koning2003NPA713} optical model calculations are shown as short-dashed red lines.}
 \label{Fig:SigRxJpnCaNi}
\end{figure}

Although the reaction cross section increases with the number of coupled inelastic states, to the limit where all open channels are coupled, Figure \ref{Fig:pNi58SigRxElab} shows that such inelastic couplings account only for a small fraction ($\approx$ 39\% at $E_{\rm lab}$ = 30 MeV) of $\sigma_{\mathrm{R}}^{\mathrm{OM}}$. However, after including couplings to the pickup channels through the CRC calculations, a large increase is found, approaching $\sigma_{\mathrm{R}}^{\mathrm{OM}}$ and the experimental data, as can be seen in Figure \ref{Fig:SigRxJpnCaNi}.
 An even better agreement can be obtained after we include the non-orthogonality terms \cite[p.\ 226]{ThompsonBookNonOrthogonality} in the CRC calculations, also shown in Figure \ref{Fig:SigRxJpnCaNi}. Non-orthogonality corrections arise because at small radii the deuteron bound state is not orthogonal to bound states occupied in the target. 
 
\begin{figure}[h]
 \begin{center}
  \includegraphics[trim = 0mm 0mm 0mm 0mm, clip,scale=0.37,angle=0.0]{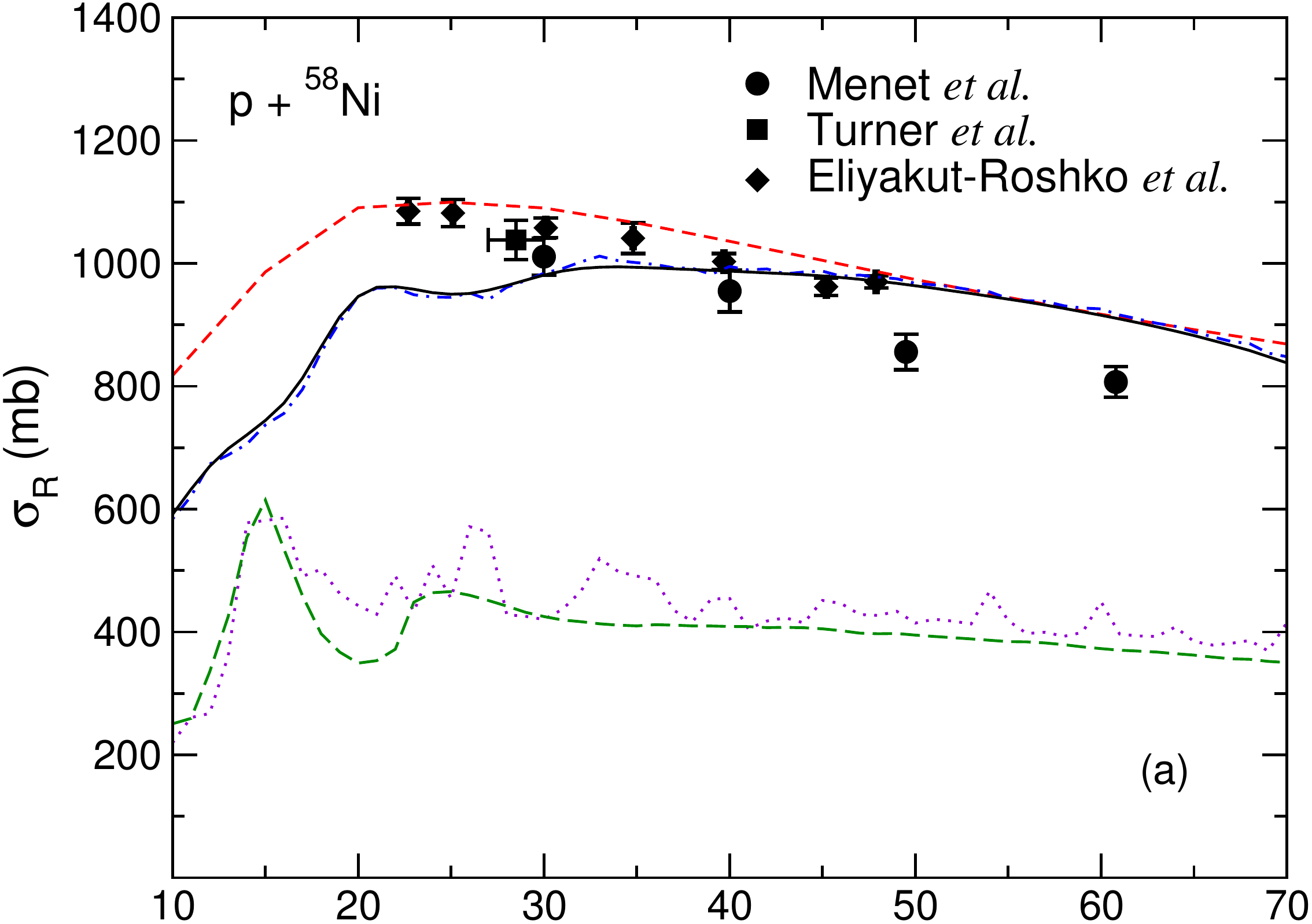} \\
  \vspace{2mm}
  \includegraphics[trim = 0mm 0mm 0mm 0mm, clip,scale=0.37,angle=0.0]{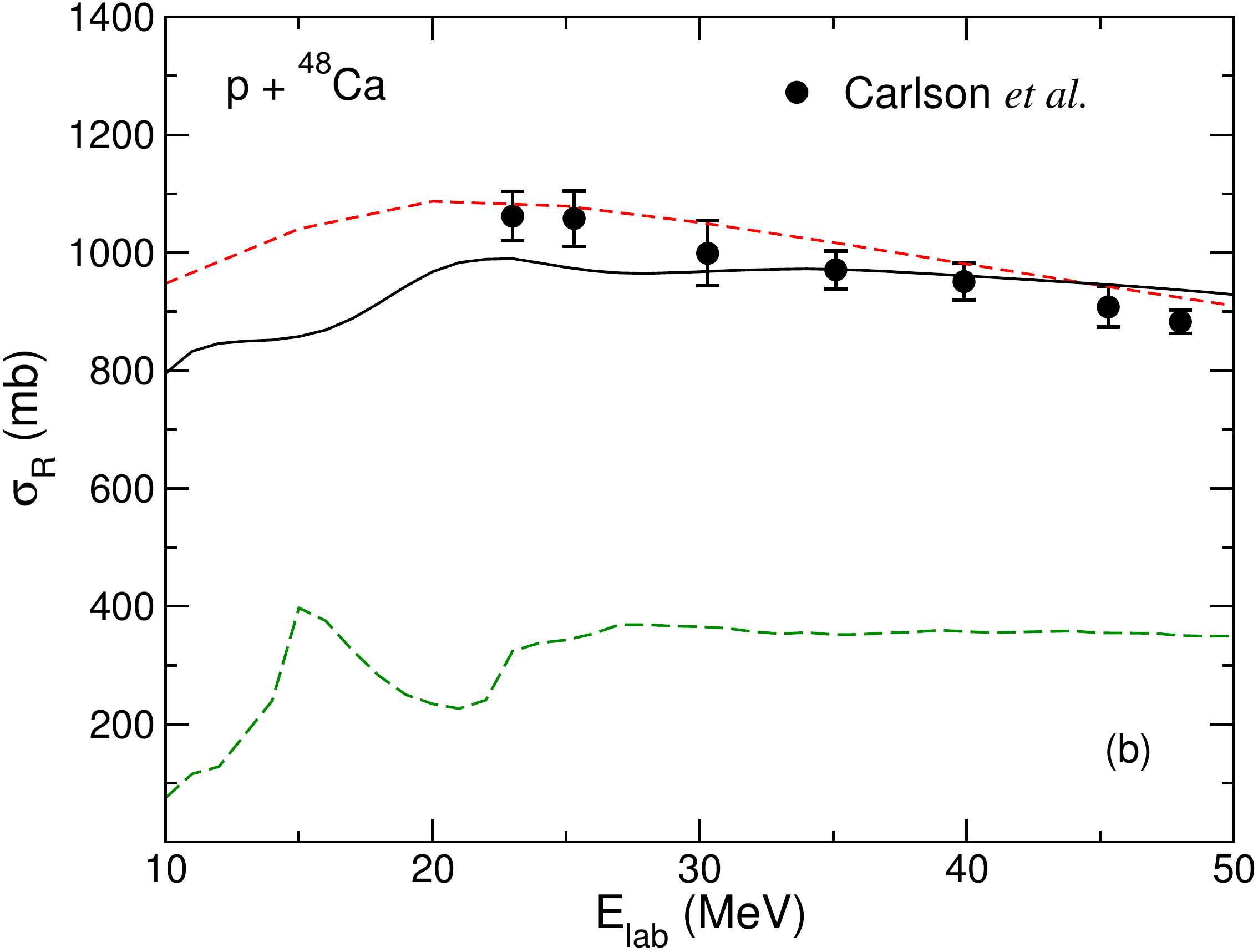} \\
 \end{center}
 \vspace{-4mm}
 \caption{(color online) Total reaction cross-section as a function of the incident energy for the reactions p + $^{58}$Ni (a) and p + $^{48}$Ca (b). The results are shown for calculations that include couplings to the damped inelastic states lying below the scattering energy (dashed green lines), and to the inelastic and transfer channels with non-orthogonality corrections (solid black lines). The Koning-Delaroche  \cite{Koning2003NPA713} optical model calculations are shown as short-dashed red lines. For the $^{58}$Ni target (a), we also show the results of two calculations that do not include damping, one for inelastic couplings (dotted magenta line) and one for inelastic and pickup channels (dash-dotted blue line). 
 Data from Refs.\ \cite{Menet1971PRC4,Turner1964NP58,Eliyakut-Roshko1995PRC51,Carlson1994PRC49}.}
 \label{Fig:pNi58Ca48Damping}
\end{figure}

In Figure \ref{Fig:pNi58Ca48Damping} we directly compare our results for p + $^{58}$Ni, $^{48}$Ca with experimental data taken from the literature. We show the reaction cross sections obtained by coupling only to inelastic open channels (i.\ e., states with $E^{*} < E_{\mathrm{lab}}$), and also by including couplings to pickup channels (with non-orthogonality corrections) in addition to the inelastic channels. As observed in Figure \ref{Fig:pNi58Ca48Damping}, despite the important role of the inelastic channels, the most significant contribution to absorption comes from the pickup channels. We achieve a good description of the experimental data after appropriately including couplings to all aforementioned processes.

Within the doorway approximation, we ignore explicitly the absorptive effects of couplings to resonances in other channels, such as compound nucleus formation. It is thus physically expected that, to compensate for the missing absorption from the neglected resonances, a damping in the non-elastic channels is necessary. In the calculations mentioned above, a damping imaginary component, corresponding to the imaginary part of the Koning-Delaroche optical potential \cite{Koning2003NPA713}, was added to the inelastic diagonal potentials. Imaginary potential components had already been included in the transfer channels. For comparison purposes, we show the result obtained when this imaginary component in the inelastic channels is not added (undamped) for p + $^{58}$Ni in the top panel of Figure \ref{Fig:pNi58Ca48Damping}. There is noticeable difference between damped and undamped calculations when including only inelastic couplings (the damped line approximately reproduces the lower limit of the undamped oscillations)
, but there is almost no discrepancy when the pickup channels are also coupled. The observed differences between the damped and undamped calculations indicate that the doorway approximation adopted, although very good, is not perfect.

\begin{figure}[h]
 \begin{center}
  \includegraphics[trim = 0mm 0mm 0mm 0mm, clip,scale=0.34,angle=0.0]{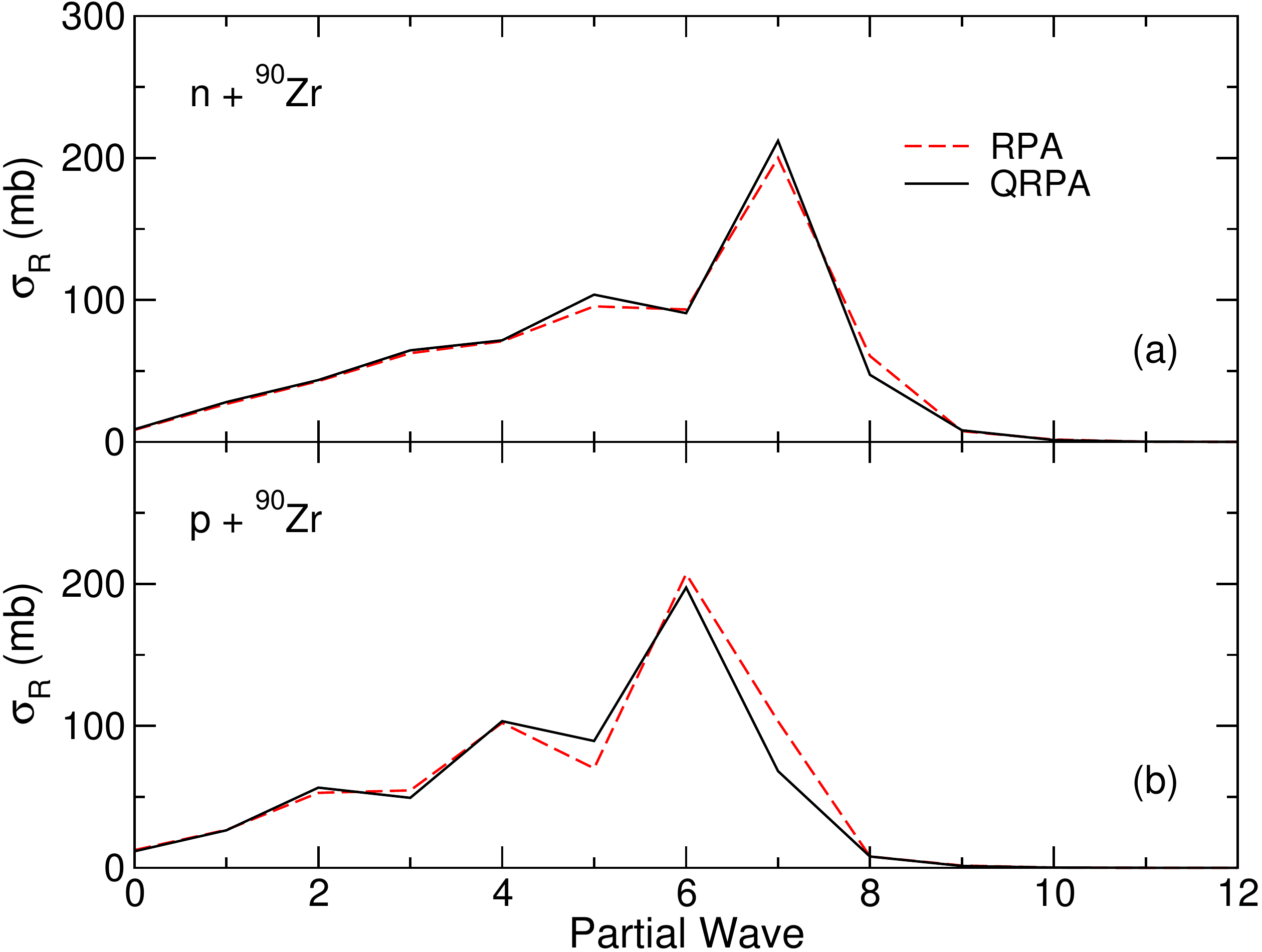} \\
 \end{center}
 \vspace{-4mm}
 \caption{(color online) Comparison of the reaction cross-section, as a function of the partial wave, between calculations using RPA states and transitions with the Gogny D1S force \cite{Decharge1980PRC21} (dashed lines) and using the QRPA model with the SLy4 functional. The results shown are for neutrons (a) and protons (b) scattered by the target $^{90}$Zr, both with $E_{\mathrm{lab}}$ = 30 MeV. Couplings to all states below 30 MeV were included, according to each model. Transfer channels were not included in the comparison.}
 \label{Fig:pZr90RPA-QRPA}
\end{figure}

\begin{figure}[h]
 \begin{center}
  \includegraphics[trim = 0mm 0mm 0mm 0mm, clip,scale=0.365,angle=0.0]{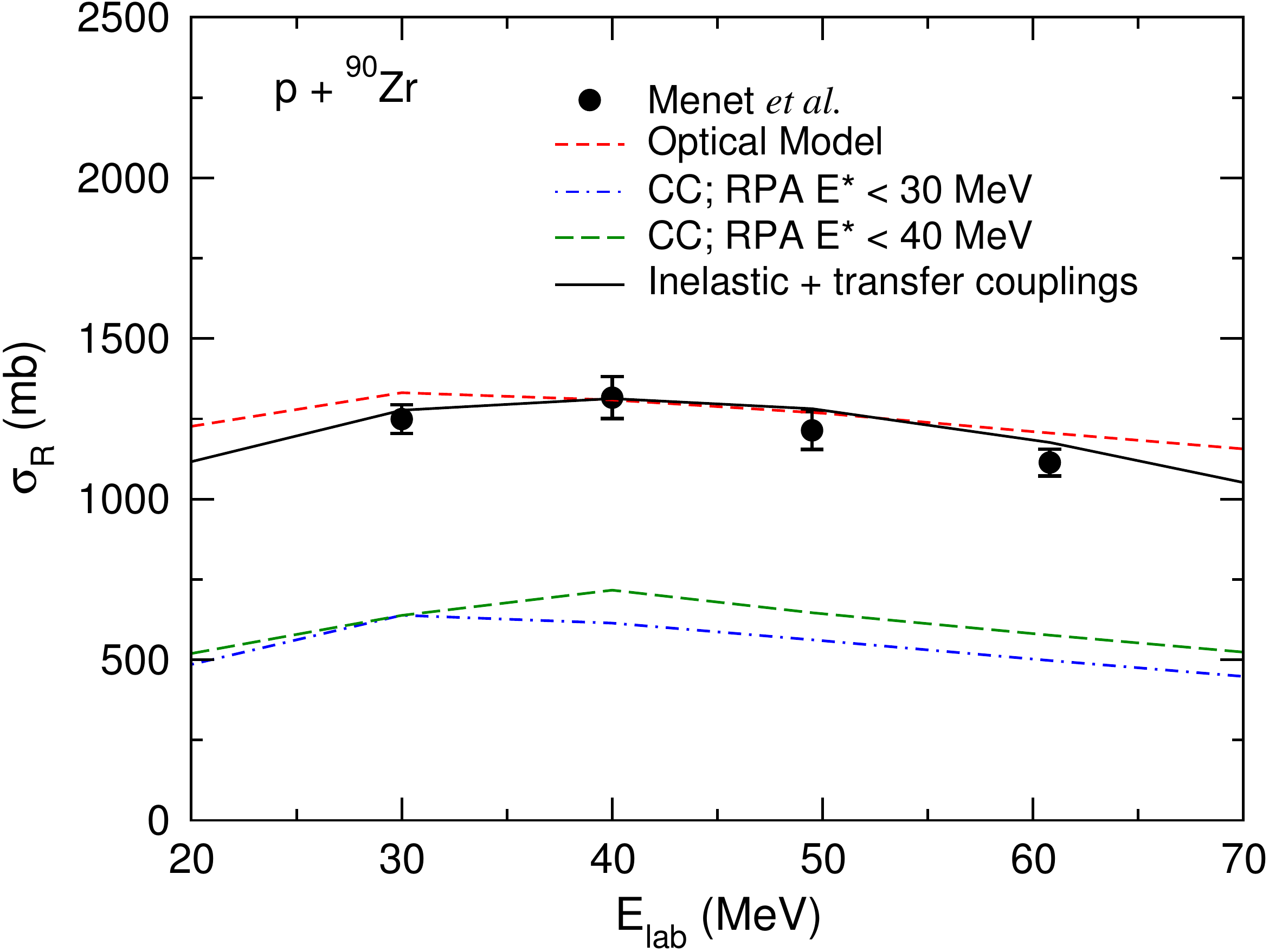} \\
 \end{center}
 \vspace{-4mm}
 \caption{(color online) Total reaction cross-section as a function of the incident energy for the reaction p + $^{90}$Zr using the Gogny D1S force \cite{Decharge1980PRC21}. The results are shown for couplings to the inelastic RPA states lying below 30 (dash-dotted line) and 40 MeV (dashed line), and to the inelastic and transfer channels with non-orthogonality corrections (solid line). The Koning-Delaroche  \cite{Koning2003NPA713} optical model calculations are shown as short-dashed lines. Data from Refs.\ \cite{Menet1971PRC4}.}
 \label{Fig:pZr90SigRxElab}
\end{figure}

A finite-range  interaction in a HFB description of the target structure was also considered, as described in Ref.\ \cite{Blaizot1977NPA284}. For reactions of nucleons scattered by $^{90}$Zr, the reaction cross section results using the QRPA model with the SLy4 functional were found to be practically equivalent to the results
found using RPA states and transitions with the Gogny D1S force \cite{Decharge1980PRC21}, as it is illustrated in Figure \ref{Fig:pZr90RPA-QRPA}. This was observed despite the proton pairing gap of 1.2 MeV of  $^{90}$Zr. The fact that, below a given excitation energy cutoff, there are generally many more QRPA states than RPA ones was not significant. In Figure \ref{Fig:pZr90SigRxElab} we show the total reaction cross sections for p + $^{90}$Zr as a function of the incident energy obtained by coupling to RPA states and using the Gogny D1S force, obtaining again good agreement with experimental data after including the pickup channels. 

\begin{figure}[h]
 \begin{center}
  \hspace{-1.2mm}\includegraphics[trim = 0mm 0mm 0mm 0mm, clip,scale=0.37,angle=0.0]{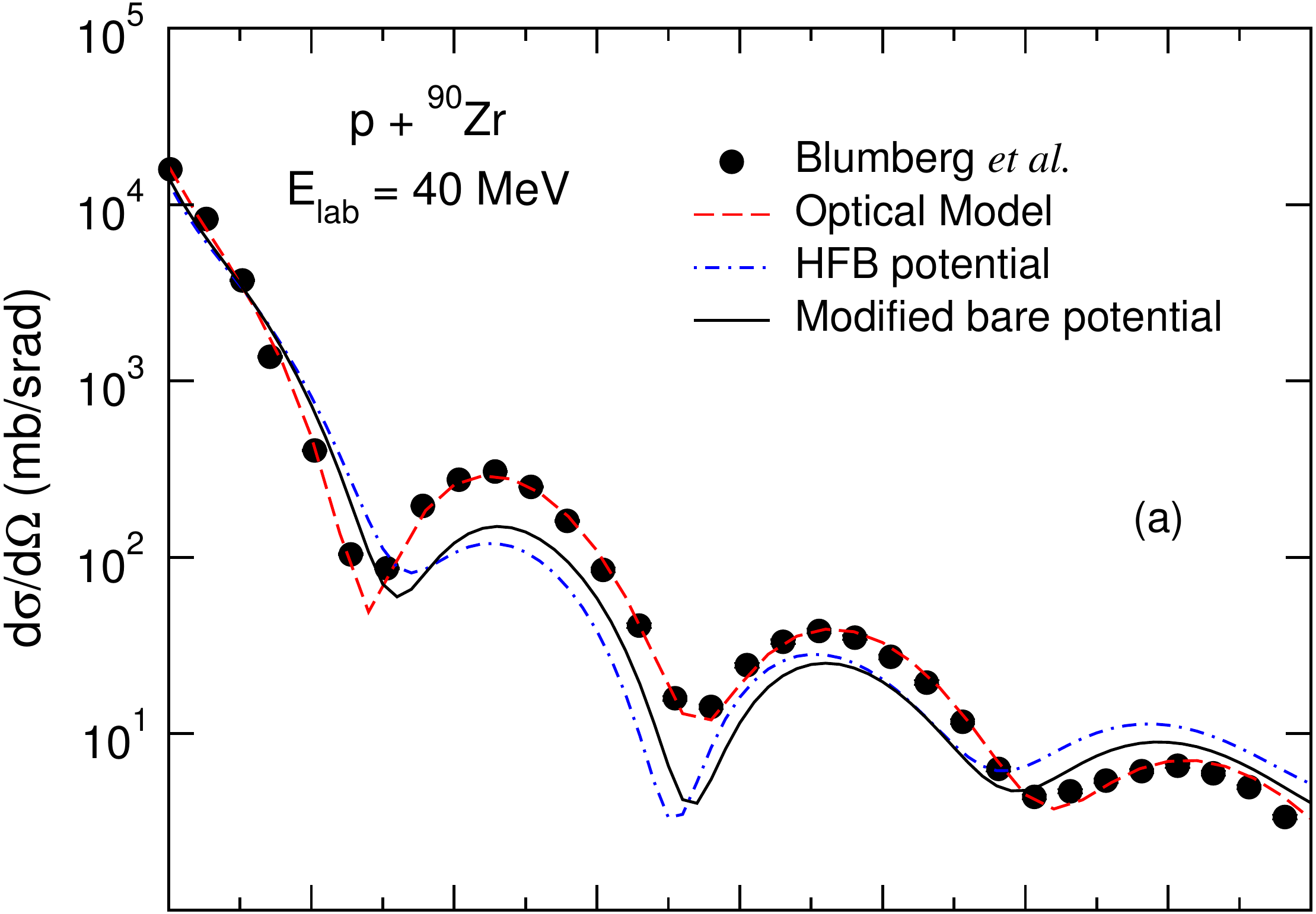} \\
\vspace{0.4mm}
  \includegraphics[trim = 0mm 0mm 0mm 0mm, clip,scale=0.37,angle=0.0]{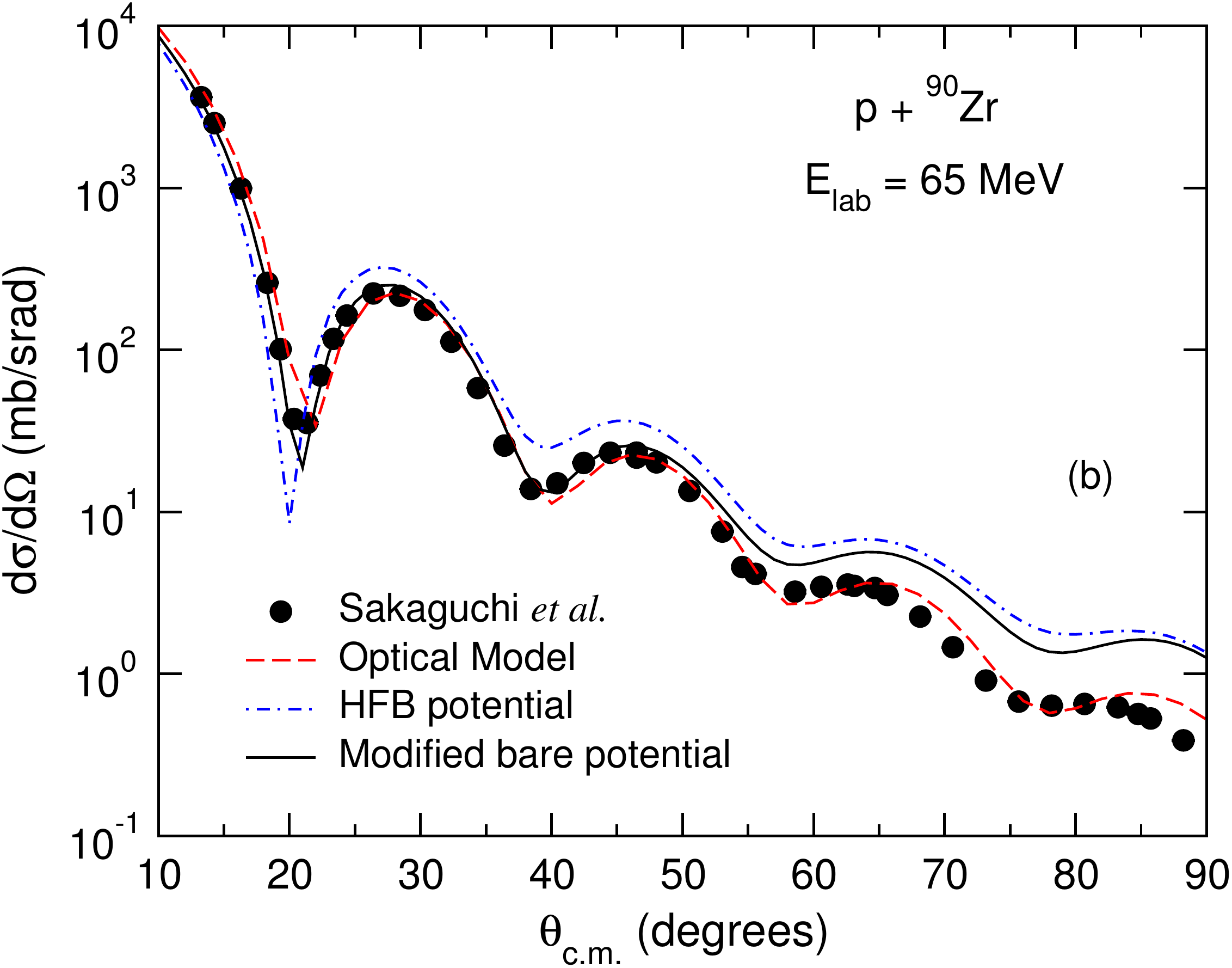} \\
 \end{center}
 \vspace{-4mm}
 \caption{(color online) Angular distribution for the elastic cross section for the reaction p + $^{90}$Zr at scattering energies of 40 (a) and 65 MeV (b). The dash-dotted and solid lines correspond to the predictions  of our model using the potential from HFB densities and from a slightly modified bare potential (see text), respectively. The Koning-Delaroche  \cite{Koning2003NPA713} optical model calculation is shown as the short-dashed line. Data from Refs.~\ \cite{Blumberg1966PR147,Sakaguchi1982PRC26}.}
 \label{Fig:pZr90AngDist}
\end{figure}

This work focuses mostly on reaction cross sections, which test the modulus of the S-matrix elements. Additional insights can be gained from elastic angular distributions. Preliminary calculations of these give reasonable agreement with measured cross sections. As an example, we show in Figure \ref{Fig:pZr90AngDist} predictions for the elastic cross sections of 40 and 65 MeV protons scattered by a $^{90}$Zr target. The dash-dotted lines represent calculations performed within our model, coupling to all QRPA inelastic and deuteron channels. A phenomenological spin-orbit component from \cite{Koning2003NPA713} was added, so the analysis of our model would be, for now, limited to the central components of the OMP. The single-folded potential obtained from the HFB ground-state density was used as the central bare potential, as described previously in Section \ref{TrPot}. The solid lines correspond to the same calculation but with a modified bare potential instead, which has a $\sim$18\% reduced internal depth while maintaining the same intensity in the surface region. This way, the internal depth of this new modified potential approaches the phenomenological one while the reaction cross sections remain essentially unchanged.
As observed in Figure \ref{Fig:pZr90AngDist}, these results are in better agreement with experimental data, especially for smaller angles, providing an additional improvement from the calculation with the unmodified potential. This illustrates how the angular distributions are sensitive to the effective interaction. This sensitivity  can serve as a method to not only evaluate how realistic the initial structure assumptions were, which were consistently extended to the calculation of reaction observables, but also to identify important features of the effective interaction adopted. For example, the fact that a better agreement with experimental data is achieved by reducing the internal depth of the bare potential indicates that density-dependence effects, which have been ignored explicitly, should be relevant for future refinements of the model. Considering the energy dependence of the interaction, or implementing an explicit treatment of the exchange terms would also improve the agreement of the model predictions with the experimental data. The result from a phenomenological optical potential \cite{Koning2003NPA713} calculation is also shown in the figure for comparative purposes.

\section{Conclusion}
\label{Sec:Conclusion}

We have calculated the reaction cross-sections for nucleon induced reactions  on nuclei    $^{40,48}$Ca, $^{58}$Ni, $^{90}$Zr and $^{144}$Sm. This was done by explicitly calculating the couplings to all relevant transfer and RPA and QRPA inelastic  channels. It was found that this inelastic convergence is achieved when all open channels are coupled. Inelastic couplings account for an important part of the reaction cross section, but still most of it corresponds to couplings to the deuteron formation pickup channel. The effect of couplings between excited levels was studied, leading to the conclusion that they did not contribute significantly to the nucleon-nucleus reaction cross sections, and thus could be ignored. We were able to obtain reaction cross sections that were in good agreement with phenomenological optical model results and experimental data, by means of coupled channels and coupled reaction channels calculations, with non-orthogonality corrections.
Absorptive effects of resonances from other channels appear to be negligible when the one-nucleon pickup channels are coupled in addition to the inelastic channels.
Preliminary calculations of elastic angular distributions were also in good agreement with experimental data, for small angles. This demonstrates the applicability of the  doorway approximation, made evident by the negligible difference in absorption observed between calculations considering damped and undamped inelastic couplings, after also including the pickup channels. Future work on couplings between different types of nonelastic processes will include higher-order corrections.

\section{Acknowledgments}
 This work was performed under the auspices of the U.S. Department of Energy by Lawrence Livermore National Laboratory under Contract DE-AC52-07NA27344, and under UNEDF collaboration with SciDAC Contract DE-FC02-07ER41457.

\appendix

\section{Fourier-Bessel Expansions}
\label{FourBess}

The following discussion is based on the treatment of Petrovich \cite{Petrovich1975NPA251}, which has been applied to inelastic scattering by Petrovich, Carr, and McManus in Ref. \cite{Petrovich1986ARNPS36}.

We consider the Fourier-Bessel series expansion of a function that can be expressed in the form
\be
f(\mathbf{r}) = f_{JM}(r) Y_{JM}^*(\mathbf{\hat{r}}),
\label{eq_fourbes_part_010}
\ee
which is defined within a radius $R$.  We refer to the radial part $f_{JM}(r)$ as a partial-wave function.  In applications the function $f_{JM}$ is frequently independent of $M$ and is then labeled $f_J$; we retain the more general form.  The Fourier transform is defined as
\be
f(\mathbf{q}) = \int d\mathbf{r} \ e^{\pm i \mathbf{q} \cdot \mathbf{r}} f(\mathbf{r}),
\label{eq_fourbes_part_020}
\ee
where the domain of the radial part of the integral is $0$ to $R$.  Petrovich employs the lower ($-$) sign; we retain both possibilities.

By using the expansion of the exponential term of Eq. (\ref{eq_fourbes_part_020})
\begin{equation}
\label{eq_fourbes_part_031}
e^{\pm i \mathbf{q} \cdot \mathbf{r}}  = 4 \pi \sum_{JM} (\pm i)^J \, j_J(qr) \ Y_{JM}^*(\mathbf{\hat{r}}) Y_{JM}(\mathbf{\hat{q}}),
\end{equation}
and the definition of the dot product of spherical tensors \cite{EdmondsAngularMomentum} and inserting the result in Eq.~(\ref{eq_fourbes_part_020}) we find:
\be
f(\mathbf{q}) = (\pm i)^J f_{JM}(q)Y_{JM}^*(\mathbf{\hat{q}}), 
\label{eq_fourbes_part_040}
\ee
where we have defined
\be
f_{JM}(q) = 4 \pi \int_0^R dr \ r^2 j_J(qr) f_{JM}(r),
\label{eq_fourbes_part_050}
\ee
which is the partial-wave form of the transformation;  we refer to the function $f_{JM}(q)$ as the partial-wave function in momentum space.

In the limit $R \rightarrow \infty$ it is easy to see that the expression for the reverse transformation is the usual Fourier integral; that is,
\be
f(\mathbf{r}) = \frac{1}{(2\pi)^3} \int d\mathbf{q} \ e^{\mp i \mathbf{q} \cdot \mathbf{r}} f(\mathbf{q}),
\label{eq_fourbes_part_100}
\ee
where the integral extends over all momentum space, or in partial-wave form
\be
f_{JM}(r) = \frac{4 \pi}{(2 \pi)^3} \int_0^\infty dq \ q^2 j_J(qr) f_{JM}(q).
\label{eq_fourbes_part_110}
\ee

If the partial wave expansion of the function to be transformed is defined in terms of $Y_{JM}$ instead of $Y_{JM}^*$, the above discussion is unaltered.  That is, if $Y_{JM}^*$ is replaced by $Y_{JM}$ in Eq.~(\ref{eq_fourbes_part_010}), all expressions are identical except for the same replacement in Eq.~(\ref{eq_fourbes_part_040}).



\bibliographystyle{unsrt}    
\bibliography{ReactionPRC}

\end{document}

%% file: comdefs.tex
\newcommand{\be}{\begin{equation}}
\newcommand{\ee}{\end{equation}}

\newcommand{\bea}{\begin{eqnarray}}
\newcommand{\eea}{\end{eqnarray}}

\newcommand{\idest}{{\it i.e.\ }}

\newcommand{\smallfrac}[2]{{\textstyle{\frac{#1}{#2}}}}

\newcommand{\signfac}[1]{(-1)^{#1}}

\newcommand{\threej}[6]
{
\left( \!\!
\begin{array}{ccc}
#1 & #2 & #3 \\
#4 & #5 & #6
\end{array} \!\!
\right)
}

\newcommand{\sixj}[6]
{
\left\{ \!\!
\begin{array}{ccc}
#1 & #2 & #3 \\
#4 & #5 & #6
\end{array} \!\!
\right\}
}

\newcommand{\ninej}[9]
{
\left\{ \!\!
\begin{array}{ccc}
#1 & #2 & #3 \\
#4 & #5 & #6 \\
#7 & #8 & #9
\end{array} \!\!
\right\}
}